# Weakly-correlated nature of ferromagnetism in non-symmorphic $CrO_2$ revealed by bulk-sensitive soft-X-ray ARPES

F. Bisti[1,2,†], V. A. Rogalev[1], M. Karolak[3], S. Paul[4], A. Gupta[4], T. Schmitt[1], G. Güntherodt[5], V. Eyert[6], G. Sangiovanni[3], G. Profeta[7,8] and V. N. Strocov[1]

[1]*Swiss Light Source, Paul Scherrer Institute, CH-5232 Villigen PSI, Switzerland*

[2]*ALBA Synchrotron Light Facility, 08290 Cerdanyola del Vallès, Spain*

[3]*Institut für Theoretische Physik und Astrophysik, Universität Würzburg, Germany*

[4]*MINT Center, University of Alabama, Tuscaloosa, Alabama 35487, USA*

[5]*II. Physikalisches Institut, RWTH Aachen University, 52074 Aachen, Germany*

[6] *Materials Design SARL, 42, Avenue Verdier, 92120 Montrouge, France*

[7]*Dipartimento di Scienze Fisiche e Chimiche, Università dell'Aquila, Via Vetoio 10, 67100, L'Aquila, Italy*

[8]*CNR-SPIN L'Aquila, Via Vetoio 10, 67100 L'Aquila, Italy*

[†] *Corresponding author: fbisti@cells.es*

## Abstract

Chromium dioxide $CrO_2$ belongs to a class of materials called ferromagnetic half-metals,

whose peculiar aspect is to act as a metal in one spin orientation and as semiconductor or insulator in the opposite one. Despite numerous experimental and theoretical studies motivated by technologically important applications of this material in spintronics, its fundamental properties such as momentum resolved electron dispersions and Fermi surface have so far remained experimentally inaccessible due to metastability of its surface that instantly reduces to amorphous $Cr_2O_3$. In this work, we demonstrate that direct access to the native electronic structure of $CrO_2$ can be achieved with soft-X-ray angle-resolved photoemission spectroscopy whose large probing depth penetrates through the $Cr_2O_3$ layer. For the first time the electronic dispersions and Fermi surface of $CrO_2$ are measured, which are fundamental prerequisites to solve the long debate on the nature of electronic correlations in this material. Since density functional theory augmented by a relatively weak local Coulomb repulsion gives an exhaustive description of our spectroscopic data, we rule out strong-coupling theories of $CrO_2$. Crucial for the correct interpretation of our experimental data in terms of the valence band dispersions is the understanding of a non-trivial spectral response of $CrO_2$ caused by interference effects in the photoemission process originating from the non-symmorphic space group of the rutile crystal structure of $CrO_2$.

**Introduction**

Among the transition metal dioxides with rutile structure, chromium dioxide ($CrO_2$) is the only one possessing a ferromagnetic conducting phase. Its ground-state Fermi surface (FS) is composed of 100% spin polarized electrons, resulting from the so-called "half-metallic" nature of $CrO_2$. For almost 30 years, the half-metallicity has been correctly predicted within density functional theory (DFT) using the local spin-density approximation (LSDA) to electron exchange-correlation [1, 2]. A clear experimental demonstration was obtained by point contact Andreev reflection, showing a spin polarization of the conductive electrons higher than 90% [3] and later in subsequent studies higher than 98% [4]. The half-metallicity of $CrO_2$ finds important practical application in spintronics. Furthermore, it was demonstrated the exciting possibility to inject a spin-triplet supercurrent into $CrO_2$ [5] which sets up interesting connections between spintronics and superconductivity. However, one of the biggest limitations to fully exploit the device potential of the half-metallicity of $CrO_2$ is its dramatic spin depolarization with temperature, which is therefore considered as a property restricted to the ground state. Several depolarization mechanisms have been suggested [6, 7], including those where electronic correlations might play an important role [7, 8].

The electron correlation effects in $CrO_2$ beyond the mean-field approach within the local density approximation are still under debate, because the experimental data reported so far diverge concerning the degree of their contribution. In the DFT-LSDA framework, the calculated density of states (DOS) does not reproduce the angle-integrated photoemission spectra of the valence band [9, 10, 11]. However, a better agreement can be obtained by introduction of an on-site Coulomb interaction term within the LSDA+$U$ approach on the Cr-$d$ orbitals using theoretically derived values of $U$ = 3 eV and $J$ = 0.87 eV [12]. On the other hand, optical conductivity [13] and magnetic anisotropy [14] are better modeled by just static LSDA; furthermore magneto-optical Kerr spectroscopy data required only the gradient corrections within the generalized gradient approximation (GGA) to the exchange-correlation [15]. $CrO_2$ has also been the subject of dynamical mean-field theory (DMFT) based calculations, including the many-body effects of the electron-electron interaction, in combination with DFT (LSDA+DMFT). In Ref. [16], the authors claimed an improved quantitative agreement in the interpretation of photoemission data with an LSDA+DMFT ($U$

= 5 eV and $J$ = 1 eV) treatment of the $t_{2g}$ orbitals as compared to LSDA or LSDA+$U$ as well as semi-quantitative agreement with thermodynamic and direct current transport measurements. A further LSDA+DMFT work [8] found the same Coulomb interaction parameters as Ref. [12] and demonstrated the appearance of non-quasiparticle states in the minority spin channel near the Fermi level ($E_F$). Their presence was claimed to be essential for the correct quantitative description of the spin polarization temperature dependence [8].

A recent comprehensive study employing static as well as dynamical treatments of correlations to investigate the mechanism leading to the peculiar half-metallic ferromagnetism in $CrO_2$ draws a complex picture [17]. According to this investigation static methods, like LSDA+$U$, were able to reproduce the magnetic ground state and some of its properties, whereas the mechanism underlying magnetic ordering could only be understood by considering dynamical correlations, inter-atomic exchange, as well as the polarization of the oxygen $p$ states. In addition, these calculations would suggest a bit smaller "standard" Slater-type Hubbard parameters: $U$ = 1.80 eV and $J$ = 0.91 eV (see Appendix: Methods for the relation between the Hubbard parameters and the so-called Kanamori parameters used by the authors, $\mathcal{U}$ = 2.84 eV and $\mathcal{J}$ =0.7 eV, to describe the interactions of the $t_{2g}$ Wannier orbitals) [17].

These results emphasize the fact that a correct treatment of electron correlations in $CrO_2$ is not just a pure theoretical issue but has important implications on its exotic transport properties. Different scenarios for the transport phenomena were suggested, depending on how the correlations were modeled [8, 13, 18]. However, a complete understanding of the nature of electron correlations in $CrO_2$ requires momentum-resolved experimental data of the electronic structure of this material. As we will show, these spectra are of paramount importance to judge the adequacy of the theoretical approach.

Angle-resolved photoemission spectroscopy (ARPES) represents a natural experimental technique to directly probe the momentum-resolved electronic structure. However, the conventional ARPES in the photon energy range 20-200 eV has a very small probing depth as characterized by photoelectron mean free path much below 1 nm. This technique is therefore inapplicable to $CrO_2$ because its surface is metastable at normal conditions and, immediately after the synthesis, develops an amorphous layer with a composition close to the antiferromagnetic $Cr_2O_3$ and a thickness of roughly 2 nm [19], well above the conventional ARPES probing depth. On the other hand, it has been demonstrated that the use of photons with higher energies towards 1200 eV [11] or lower energies towards 8 eV [20] deliver a probing depth around 2 nm, which gives the possibility to detect the photoelectrons emitted from $CrO_2$ through the $Cr_2O_3$ overlayer.

In this work, we demonstrate that the momentum-resolved electron dispersions and FS of $CrO_2$ can be explored by bulk sensitive soft-X-ray ARPES (SX-ARPES) using photon energies in the range of 320-820 eV. Moreover, the increase of the photoelectron mean free path in this energy range reduces, by the Heisenberg relation, the intrinsic uncertainty of the momentum $k_z$ perpendicular to the surface [21], allowing an accurate mapping of the 3D electron dispersions. We find that the experimental FS appears as composed mainly of an electron pocket around the Γ point and a hole pocket around the Z point of the tetragonal Brillouin zone. Our first-principles DFT calculations, taking into account non-trivial matrix element effects in the ARPES response of $CrO_2$ originating from its non-symmorphic rutile space group, demonstrate that the spin-polarized GGA+$U$ approximation (in what follows GGA implies an explicitly spin-polarized functional also in abbreviations such as GGA+$U$ and GGA+DMFT unless stated otherwise) with the on-site parameter $U_{eff}$ ($U_{eff} = U - J$) [22] equal to 0.4 eV (1 eV with the use of LSDA) delivers an accurate description of the $CrO_2$ band structure. Additional calculations within the DFT+DMFT framework allowed us to clarify this point, by showing almost no modification with respect to the DFT+$U$ band structure. We conclude therefore that the occupied band structure of $CrO_2$, below the magnetic ordering transition temperature, is compatible with a rather weakly correlated scenario, with electron correlation effects being essentially exhausted by static mean-field theory.

**Results**

**Fermi surface and photoemission interference effects**

In Fig. 1 we gather our theoretical and experimental information on the FS of $CrO_2$. The FS obtained from our GGA+$U$ calculations with $U_{eff}$ = 0.4 eV (for the determination of $U_{eff}$ see below) is reported in Fig. 1(*a*) inscribed into the first Brillouin zone (BZ). The theoretical FS is fully spin polarized and characterized by a quasi-isotropic electron pocket around the Γ point (violet surface), a hole pocket along the Γ-Z direction (yellow) that barely closes near the Z points, and another electron pocket around the A point (violet). Different colored planes *p*1-*p*4 in Fig. 1(*a*) show FS cuts explored in our SX-ARPES experiment, the surface-parallel cut *p*1 under sample rotation and the surface-perpendicular cuts *p*2-*p*4 under variation of photon energy. The top maps in the respective panels *p*1-*p*4 report the corresponding ARPES intensity rendered into the electron momentum coordinates $k_x$, $k_y$ and $k_z$ corrected for the incident X-ray photon momentum. The sharpness of the FS contours in the $k_z$ direction confirms sharp definition of $k_z$ resulting from the photoelectron mean free path increase in the SX-ARPES energy range [21].

The bottom maps in the *p*1-*p*4 panels in Fig. 1 report the same experimental data overlaid with the FS contours obtained from GGA+$U$ calculations. The contours shown on the $k_{x,y}>0$ side of these maps (marked "2-Cr BZ") correspond to the full rutile unit cell of $CrO_2$ including two Cr atoms, and the ones on the $k_{x,y}<0$ side (marked "1-Cr BZ") unfold these contours onto a reduced body-centered tetragonal unit cell including one Cr atom (see below). The most striking feature of our data, in comparison with the full unit cell calculations, is particularly evident in the $p2$ cut: the electron pocket around Γ (violet contours) is experimentally present only in every second BZ, centered at the surface-perpendicular momenta $k_z$ equal to even integers $n$ of $2\pi/a$ (in our case $n = 8$ and 10), and disappears in those centered at odd integers ($n = 7$ and 9 designated as Γ'). Complementarily, in the next BZ along the surface-parallel momentum $k_x$ represented in the cut $p3$, the electron pockets (around Γ, violet contours) are visible at $k_z$ equal to the even integers of $2\pi/a$ but disappear at odd ones. Furthermore, in the $p2$ cut we distinguish also the hole pockets (yellow contours) which show the same odd-even alternation. In this way, our data exhibit a periodicity in reciprocal space larger than expected, meaning that in real space this periodicity should be related to an effective unit cell smaller than the nominal one. Similar mismatch between the real space periodicity and that obtained in ARPES spectra has recently been observed in iron pnictides [23, 24, 25] as well as in a series of materials whose crystal structure possesses non-symmorphic space group such as graphite [26], BiTeCl [27], decagonal Al–Ni–Co quasicrystal [28] and Ruddlesden-Popper iridates [29].

In the case of $CrO_2$ the apparent twice as large periodicity of the ARPES response in **k**-space originates from the fact that its rutile-type space group ($D^{14}_{4h}$: $P4_2/mnm$) is non-symmorphic. The Cr atoms form a body-centered tetragonal lattice but are surrounded by distorted octahedra of oxygen atoms (see Fig. 1b) with the octahedron in the center being rotated by 90° around the $c$-axis with respect to the octahedra at the corners of the tetragonal cell. The presence of this screw axis reduces the accessible final states in the photoemission process belonging to different irreducible representations (even or odd with respect to the screw axis symmetry) along the high symmetry lines, which appears as alternating visibility of the spectral features through the successive BZs [26, 30]. For reproducing this selection rule in the photoemission process, a possible way is to perform a formal unfolding of the band structure from the full to an effective unit cell, as has recently been demonstrated for iron pnictides [23] and for the Weyl semimetal WTe$_2$ [31]. Briefly, this procedure is based on projection of the Bloch wave functions onto a basis set that is just even or odd with respect to the screw axis symmetry. In our case, a simple basis set is the one suggested by the crystal field theory, *i.e.* the $d_{xy}$, $d_{yz-zx}$ and $d_{yz+zx}$ atomic orbitals of the two different Cr atoms in their local coordinate frame with respect to the oxygen octahedron (for further details see 'Unfolding procedure' in the Appendix: Methods). In practice, the unfolding procedure assigns to the first-principles eigenvalues weights proportional to the projections of the corresponding eigenfunctions onto the new basis set. The results of this procedure performed with our GGA+$U$ calculations, on the basis set even with respect to the screw axis symmetry (see 'Unfolding procedure' in the Appendix: Methods), are reported in the bottom maps of Fig. 1 on their left ($k_{x,y}$<0) side marked "1-Cr BZ". The unfolding perfectly reproduces the experimental odd-even visibility of the electron pockets around the Γ and Γ' points in *p*1-*p*2 as well as the hole pockets around the Z point in *p*1.

Having understood this crucial aspect of the ARPES response of CrO$_2$ we will discuss other FS cuts in Fig. 1. Due to the unit cell symmetry between the $k_x$ and $k_z$ directions, the ($k_y$, $k_x$) cut in the panel *p*1 is identical to the ($k_y$, $k_z$) one in the panel *p*2, although the latter shows the FS contours clearly affected by the intrinsic $k_z$ broadening [21]. Furthermore, the

*p*4 panel clearly reveals the small electron pocket around the A point in perfect agreement with the GGA+*U* predictions, Fig. 1 (a).

Sharp contrast and excellent statistics of our SX-ARPES data confirm that this technique is indeed capable of digging out the electronic structure of $CrO_2$ through the $Cr_2O_3$ overlayer. The experiment reveals the FS topology as composed of two electron pockets around the Γ and A points and one hole pocket between the Γ and Z points, Fig. 1(*a*). Taking into account the matrix element effects, the experimental results are fully consistent with our GGA+*U* calculations where all Fermi states belong to the majority spin channel and have the $d_{yz-zx}$ and $d_{yz+zx}$ character. The agreement extends to both topology and Luttinger volume of the FS pockets.

**Band dispersions**

Further information about the electronic structure of $CrO_2$ is contained in the experimental band dispersions along the Γ-X and Γ-Z directions reported in Fig. 2. Each row of the panels *a-d*, from left to right first shows the raw ARPES image. On top of the dispersive coherent spectral component, each image contains a large non-dispersive component centered around -1.5 eV, which can be modeled by angle integration of the raw images, as shown in the next panel. This component is formed, first, by photoexcitation in the amorphous $Cr_2O_3$ overlayer and, second, by photoelectrons excited in $CrO_2$ and quasielastically scattered in $Cr_2O_3$ on their escape to vacuum and thus reflecting the **k**-integrated DOS of $CrO_2$. However, we note that the experimental non-dispersive component is dominated by photoexcitation in amorphous $Cr_2O_3$ because it strongly deviates from the expected DOS of $CrO_2$ in Fig. 3 (*a*). According to previous work by Li *et al.* [32] the $Cr_2O_3$ valence band spectrum is indeed characterized by a broad localized peak at -1.5 eV (with full width at half maximum of about 1.2 eV) and other peaks starting from -5 eV and extending to lower binding energy. This fact calls for re-consideration of the previous works, where the value of $U_{eff}$ was estimated by comparison with angle-integrated spectra. The coherent component characteristic of the crystalline $CrO_2$ is obtained by subtracting the non-dispersive component from the raw images (see 'Data Processing' in the Appendix: Methods) as shown in the central panels of each row of Fig. 2. The extracted coherent components reveal sharp band dispersions, allowing direct comparison with our GGA+$U$ calculations on the rightmost panels in Fig. 2. Overlaid on the same experimental data, the unfolded theoretical bands presented on the $k_x$<0 side (marked "1-Cr BZ") again truly reproduce the matrix element effects in our data in comparison with the ones without unfolding shown on the $k_x$>0 side (marked "2-Cr BZ"). In particular, the calculation along the Γ-X direction ($k_x$<0) correctly reproduces the $d_{yz+zx}$-like band forming the central FS pocket around the Γ point (although slightly overestimates its size) and the unfolding reproduces the cancellation of this band in the second BZ. The same striking agreement between experiment and the unfolded calculated bands is found along the Γ-Z direction (except the deep oxygen-derived $sp$-states which are not correctly described by the $d$-orbitals of our unfolding basis set).

The dispersions along the Γ-Z direction are most sensitive to variations of $U_{eff}$ in our GGA+$U$ calculations (see below) and were therefore carefully investigated using different photon energies and polarizations of incident X-rays. A zoom-in of the near-$E_F$ region measured at 603 eV ($k_z = 9\cdot 2\pi/a$) is reported in Fig. 2 (*c*) and linear dichroism at 748 eV ($k_z = 10\cdot 2\pi/a$) in (*b*, *d*). Changing the incoming light polarization from *p*- to *s*-polarization switches the ARPES response between different sets of bands (even or odd with respect to the screw axis symmetry operator). The same linear dichroism has been noted in ARPES experiments on $TiO_2$ and interpreted as switching from Γ to the next Γ' point [30]. We note that the band calculations without unfolding ("2-Cr BZ") demonstrate the absence of hybridization between different bands in their intersections at the X and Z points (in particular, in the cone at the Z point near $E_F$). This effect appears because the two intersecting bands belong to different irreducible representations (even or odd) of the non-symmorphic space group $D^{14}_{4h}$:$P4_2/mnm$ of $CrO_2$. Naturally, in the unfolded band representation ("1-Cr BZ") the weight of one of the two branches vanishes (see also Fig. 5 in the Appendix: Methods).

**Determination of the static effective Coulomb interaction**

For comparison with experiment, we have used the simplified approach by Dudarev *et al.* [22] (DFT+$U_{eff}$) since its one-parameter form facilitates fitting with the data. The effect of the on-site Coulomb interaction $U_{eff}$ on the majority spin band structure calculated within the GGA+$U$ scheme is reported in Fig. 3(*b*). Higher $U_{eff}$ values push the $d_{xy}$ band to higher binding energy (most visible along the Γ-Z direction, as noted before, in the interval from -0.6 to -1.1 eV) and, at the same time, push the $d_{yz+zx}$ band closer to $E_F$. The most satisfactory match between the GGA+$U$ calculated and experimental data (grey points obtained by fitting of the spectral peaks) is achieved as the best compromise between these two trends reached with $U_{eff}$ = 0.4 eV. Furthermore, in Fig. 3(b) we also report the calculations using LSDA for the static exchange-correlation. The same good agreement with experiment is achieved, but in this case with larger $U_{eff}$ = 1 eV (the difference is attributed mainly to higher accuracy of the equilibrium lattice in GGA than in LSDA [33, 34]). The effect of $U_{eff}$ on the FS contours is reported in Fig. 3(*c*). Their modifications with $U_{eff}$ are essentially restricted by the neighborhood of the Z and A points since the on-site Coulomb interaction do not much affect energies of the $d_{yz+zx}$ and $d_{yz-zx}$ orbitals.

Such small values of $U$ were used before to describe the magnetocrystalline anisotropy of $CrO_2$ [14], where the authors concluded that correlation effects might be important although they are strongly screened out. A full description of screening mechanisms [35, 36, 37, 38] in $CrO_2$ is particularly complicated because of the presence of many different screening channels (*d-d*, *p-d* and others). In addition to the screening channels, the choice of orbital basis and the corresponding polarization processes is another crucial point [39]. A complete discussion of the screening in $CrO_2$ is beyond the scope of this work.

For completeness, we complemented the simplified approach by Dudarev *et al.* [22] with calculations using the approach by Liechtenstein *et al.* [40], which includes the full matrix structure of the atomic Coulomb interaction. In this case, the same good agreement with experiment can be achieved within GGA+$U$ using $U$ = 1.0 eV and $J$ = 0.87 eV.

These rather smaller interaction values with respect to the recently derived ones from first principles in Ref. [17] find a reasonable explanation by the presence of a large exchange splitting [14] already in the spin-polarized exchange-correlation functional used here. To illustrate this important point, we have performed an additional GGA+$U$ calculation within Liechtenstein's scheme utilizing a spin-averaged GGA part ($S_{AVG}$-GGA). Here, the calculation as a whole is spin-polarized, however, the GGA exchange correlation functional is spin-averaged and the spin polarization stems only from the Coulomb interaction. This is in the spirit of the usual DFT+DMFT scheme, where one starts from a spin-degenerate DFT calculation. Using this approach we obtain $U$ = 2.0 eV and $J$ = 0.87 eV, which are remarkably close to the derived ones [17].

Beside the specific values of the static effective Coulomb interaction, the comparison between theory and the measured **k**-resolved photoemission data for $CrO_2$ clearly indicates that a static mean-field method gives a very good description.

**Dynamical mean-field theory investigation**

The electronic band structure has also been investigated within the DFT+DMFT framework (see Appendix: Methods for details). We used a $t_{2g}$ model derived once from $S_{AVG}$GGA calculations and also from GGA calculations, with the combination of $U$ = 2.0 eV, $J$ = 0.87 eV and $U$ = 1.0 eV, $J$ = 0.87 eV, respectively, as discussed before. In this way we can explicitly track the influence of the dynamical treatment of the interaction.

The results are summarized in Fig. 4 showing a poor agreement of $S_{AVG}$GGA+DMFT calculations, but instead a very good one with GGA+DMFT. We also have checked that the agreement with experiment for DFT+DMFT does not improve substantially by extending the model to the full 3d shell of Cr. In addition, since the GGA+DMFT calculation agrees very well with the GGA+$U$ results, we have confirmed that static mean-field theories are sufficient to describe the magnetically-ordered phase of $CrO_2$. The absence of additional effects of dynamical correlations allows us to consider $CrO_2$ as weakly correlated in this sense.

The particular intensity reduction of the lower band for the $d_{yz+xz}$ orbital is rooted in the dynamical (frequency dependent) self-energy of the latter. Within GGA+$U$ the static self-energy is a purely real quantity and thus can only lead to shifts of the bands, while DMFT includes both the real and imaginary parts of the self-energy. This difference is crucial, since the imaginary part is responsible for an energy dependent broadening of spectral features and is thus responsible for the observed changes in the spectrum of the $d_{yz+xz}$ orbital. The stronger Coulomb interaction needed within $S_{AVG}$GGA+$U$ and $S_{AVG}$GGA+DMFT is ultimately responsible for the qualitative difference between the **k**-resolved spectra in the static and dynamic approximations. Within the GGA+$U$ and GGA+DMFT approaches the Coulomb interaction is instead small enough for the differences not to matter yet. We have explicitly tested that increasing $U$ in GGA+DMFT to the $S_{AVG}$GGA+DMFT value suppresses the lower branch of the $d_{yz+xz}$ band. On the other hand, a mere reduction of the $U$ value does not solve the problems of $S_{AVG}$GGA+DMFT enforced at the GGA level. We have tried different values of $U$ between 1 eV and 2 eV, finding that at about 1.6 eV the lower branch of the $d_{yz+xz}$ orbital is recovered. At this value the $d_{xy}$ and $d_{yz-xz}$ orbitals almost touch the Fermi level along Γ-Z and X-Γ, respectively, in disagreement with ARPES. Also the complete spin polarization of the material is slowly lost upon reduction of $U$ with minority spin spectral features of $d_{yz\pm xz}$ character appearing at the Fermi level. We speculate that these are the non-quasiparticle states as seen in the variational cluster approach (VCA) and DMFT calculations of Ref. [8]. Our spectrum in fact becomes similar to that of VCA with the same poor agreement with ARPES if we make the Coulomb interaction larger [41]. However, the data by Huang et al. [42] suggesting a rapid drop of the spin polarization above the Fermi level would be better explained in $S_{AVG}$GGA+DMFT. It thus appears that the "dualistic electronic nature" of $CrO_2$, a phrase coined in Ref. [42], is also present in our calculations in the sense that the occupied electronic structure measured in ARPES is very well reproduced within GGA+$U$ and GGA+DMFT with small Coulomb interaction. However, the unoccupied electronic structure reported in Ref. [42] is better modeled in $S_{AVG}$GGA+DMFT. A unified description capable of describing both occupied and unoccupied states is apparently beyond the $t_{2g}$ model used here.

There are also other theoretical points that have to be considered, also touched upon in Ref. [17]. Since the approximations made to the Coulomb interaction are not detrimental to the agreement of the **k**-resolved spectrum of magnetic $CrO_2$ with experiment as confirmed by GGA+DMFT falling on top of GGA+$U$ one could infer that the issue is actually rooted somewhere else. The GGA+$U$ calculations that successfully describe $CrO_2$ are all-electron charge self-consistent. This means that the dynamical hybridization between the Cr 3$d$ and the O 2$p$ states including rearrangements of the charge between atoms/orbitals is fully taken into account. A recent theoretical paper on the $CrO_2$ paramagnetic phase electronic structure indicates that, under the DFT+DMFT framework, there is an important hybridization effect of the oxygen $p$ states with the chromium $t_{2g}$ [43]. In the ferromagnetic state the issue of the hybridization is even more complicated. The majority $t_{2g}$ bands overlap with the O 2$p$ bands, on the other hand the minority $t_{2g}$ bands are separated from O 2$p$ by a gap of > 1eV. This leads to different hybridization strengths in the two $t_{2g}$ spin channels, an effect that appears to be crucial for the correct description of the occupied spectrum. In $S_{AVG}$GGA+DMFT the hybridizations are by construction identical on the GGA level and become different within DMFT only after inclusion of the Coulomb interaction as the system polarizes. However, the Coulomb interaction alone, cannot mimic the different hybridizations produced by a charge self-consistent spin-polarized GGA and thus the results of GGA+DMFT and $S_{AVG}$GGA+DMFT differ strongly.

To remedy the unexpectedly bad performance of the $S_{AVG}$GGA+DMFT as compared to ARPES, one has to go beyond the minimal $t_{2g}$ model used here. An explicit inclusion of the O 2$p$ states into the DMFT loop (ideally taking care of local $p$-$p$ interactions $U_{pp}$ as well as interatomic $U_{pd}$) as well as a charge feedback into the DFT part to capture changes in the dynamical hybridization would probably be necessary in this case.

Another point that finally could be of importance is the momentum dependence of the self-energy, which is neglected hitherto. These additional non-local terms could lead to a reduced impact of the purely local correlation effects and thus to an improvement in the description of the $d_{yz+xz}$ spectrum also within an approach based on a non-spin-polarized one-particle starting guess such as *GW*.

**Discussion**

The crucial element of our exhaustive picture of the electronic structure of the paradigm half-metallic ferromagnet CrO$_2$ has been the use of SX-ARPES providing large probing depth and sharp definition of three-dimensional electron momentum. The non-trivial interference effects in the ARPES response of CrO$_2$ caused by its non-symmorphic space group were described by band structure unfolding onto an effective smaller unit cell using atomic projections. Deciphered with the unfolding, our experimental dataset yields the most complete momentum-resolved description of the electron dispersions and Fermi surface so far achieved for CrO$_2$. We establish its FS topology as composed of two electron pockets around the Γ and A point and a hole pocket along the Γ-Z direction. At least at low temperature, our theoretical description finds full spin polarization of the states at and close to the Fermi energy confirming the half-metallic nature of CrO$_2$. The resulting knowledge of the orbital character of the bands reveals that in the itinerant ferromagnetic phase of CrO$_2$ the electronic states near the Fermi energy are composed of the three-dimensionally dispersing $d_{yz+zx}$ and $d_{yz-zx}$ orbitals, and the more localized rather one-dimensional $d_{xy}$ orbital, which mediates metal-metal overlap parallel to the rutile *c*-axis (see Ref. [44] for a more detailed discussion of the orbitals).

Excellent agreement of our GGA+$U_{eff}$ calculations with a relatively small value of $U_{eff}$ (especially when combined with the GGA exchange-correlation functional) evidences that CrO$_2$ is a weakly correlated magnet where the electron correlations are essentially exhausted by spin-polarized mean-field theory plus a small correction obtained using a simple parametrization of the Coulomb interaction. This finding evidences excellent support from our DMFT calculations and is in agreement with previous work by Mazin *et al.* as well as by Toropova *et al.* [13, 14].

The picture of weak electron correlations in the magnetically ordered phase of $CrO_2$ has important implications on the debated mechanism of its depolarization at finite temperature [6, 7]. Indeed, this picture would support depolarization induced by single-electron mechanisms such as formation of sublattices with different spin or phonon interactions rather than electron correlation effects such as non-quasiparticle or orbital Kondo interactions [6, 7, 8]. However, it is important to note that our data are only on the occupied part of the electronic structure, therefore we cannot exclude the alternative scenario of a "dualistic electronic nature" [42], where the occupied electronic structure contains weak correlation and the unoccupied electronic structure involves stronger ones.

Since the GGA+*U* calculations are well capable of reproducing the experimental data, they serve as a good starting point for discussing the half-metallic ferromagnetic behavior. Obviously, $CrO_2$ eventually ends up in a half-metallic state because the exchange splitting of the $t_{2g}$ bands (of about 3 eV) is considerably larger than the widths of the $d_{yz+zx}$ and $d_{yz-zx}$ bands (of about 2.5 and 2 eV for the spin majority and minority bands, respectively), which themselves are wider than the $d_{xy}$ band (of about 1.5 eV width). In this context, it is important to point to the small hybridization between the one-dimensional $d_{xy}$ bands and the three-dimensionally dispersing $d_{yz+zx}$ and $d_{yz-zx}$ bands as has been clearly revealed by orbital-weighted band structures for $CrO_2$ [14] and the neighboring rutile-type dioxides [44, 45, 46]. Consequently, these two types of bands (the 1D $d_{xy}$ and the 3D $d_{yz+zx}$ and $d_{yz-zx}$) disperse rather independently from each other and are coupled only via the common Fermi energy, *i.e.* by charge balance. The positions and widths of the $d_{yz+zx}$ and $d_{yz-zx}$ bands are rather determined by the π-type overlap of the $t_{2g}$ orbitals with the O p states. In contrast, the $d_{xy}$ orbitals are particularly susceptible to both strong metal-metal overlap and strong (magnetic) correlations, which have strong impact on the band formation. In $CrO_2$, the peculiar localized character of these bands causes a high density of states close to the Fermi level (however without participating in the Fermi surface itself) and thus is of fundamental importance for the ferromagnetic stabilization. In contrast, isoelectronic $MoO_2$, rather than displaying a ferromagnetic phase, experiences a distinct lattice distortion away from the rutile structure to a monoclinic structure [46]. The different deviations of the two $d^2$ oxides away from a non-magnetic rutile-type ground state are thus reasonably related to the different band widths of the respective d states. Eventually, we are thus left to conclude that the half-metallic-behavior is mostly determined by the intra-atomic exchange coupling.

**Conclusions and perspectives**

Our study bears a number of new perspectives on the spectroscopic and theoretical side: (1) We have illustrated the spectroscopic power of SX-ARPES whose enhanced probing depth gives the access to **k**-resolved electronic structure of materials, where intrinsic perturbations of crystallinity or stoichiometry in the surface region (like reduction of $CrO_2$ to $Cr_2O_3$ in our case) hinder application of conventional VUV-ARPES. This methodology extends to an overwhelming variety of materials whose atomically clean surface cannot be prepared in-situ and to samples transferred ex-situ [47]; (2) On the theoretical side, we have demonstrated the utility of the unfolding method for correct interpretation of ARPES data from systems with non-symmorphic crystal structures. The recent outburst of interest to the non-symmorphic structures has been due to their crucial role in materials with strong spin-orbit coupling, where they protect band crossing forming new topological states of condensed matter such as the massless Dirac and Weyl fermions. Examples of such states include Dirac, Weyl and node-line semimetals (for entries see, for example, [48] and the references therein), nodal-chain metals [49] and organic metals [50]. Moreover, the sharp $k_z$ definition achieved in SX-ARPES due to enhanced probing depth allows accurate resolution of the topological band configurations in 3D **k**-space such as the Weyl cones [51, 52, 53]. The fundamental scientific and application perspective of the topological systems has been recognized by the Nobel prize 2016. (3) Our study puts a new perspective to utilize the robust half-metallicity of $CrO_2$ arising from its weakly correlated nature to engineer new quantum phases in $CrO_2$ based heterostructures. For example, interfaces of $CrO_2$ with $TiO_2$ are theoretically predicted to host a single-spin Dirac fermion phase which, in superlattice configuration, transforms under spin-orbit coupling into a Chern insulator phase exhibiting quantum anomalous Hall effect [54]. Another example is a heterostructure of thin half-metal $CrO_2$ layers with *s*-wave superconductors that may form topological superconductivity with Majorana fermion edge state [55] bearing a potential for topological quantum computations [56]. Finally, the proximity of $CrO_2$ nanowires to superconductors may form Josephson junctions carrying long-range supercurrents for dissipationless spintronics [57].

## Acknowledgements

We thank H. Dil, J. Minar, K. Held, Yu. S. Dedkov, L. Chioncel, and O. Gunnarsson for promoting discussions. The research leading to these results has received funding from the Swiss National Science Foundation under the grant agreement n.° 200021_146890 and European Community's Seventh Framework Programme (FP7/2007-2013) under the grant agreement n.°290605 (PSI-FELLOW/COFUND).

The authors gratefully acknowledge the Gauss Centre for Supercomputing e.V. (www.gauss-

centre.eu) for funding this project by providing computing time on the GCS Supercomputer SuperMUC at Leibniz Supercomputing Centre (LRZ, www.lrz.de).

This work was supported by the DFG through FOR 1162 (M.K.) and SFB 1170 'ToCoTronics' (G.S.).

## Appendix: Methods

**Samples and SX-ARPES measurements**

Epitaxial thin films of $CrO_2$ (100) were grown by chemical vapor deposition in oxygen atmosphere on top of a $TiO_2$ (100) substrate [58] at the MINT Center, University of Alabama, USA. The SX-ARPES experiments at different polarizations of incident X-rays were performed at the SX-ARPES endstation [59] of the ADRESS beamline [60] at the Swiss Light Source synchrotron facility, Villigen-PSI, Switzerland. The samples were transferred for SX-ARPES measurements ex-situ without any treatment. The sample temperature during the measurements was around 12 K to quench suppression of the coherent spectral weight due to thermal effects [61]. The combined (beamline and analyzer) energy resolution was set to vary between 40 and 100 meV through the incident photon energy range 300-900 eV. The sample surface was oriented normal to the analyzer axis, and the grazing incidence angle of photons was 20°. Details of the experimental geometry as well as photon momentum corrected transformation of the emission angles and energies into **k** values can be found in [59].

**Data processing**

The experimental FS maps in Fig. 1 were obtained by the integration of the spectral intensity within ±0.05 eV around $E_F$. The data of each slice composing the FS maps were normalized to the integral intensity over 90% of their angular range, this allows to compensate the photoexcitation cross-section variation over the large photon energy range of panels *p2*, *p3* and *p4*, and the slight variation of the probed region in the panel *p1* due to the angular scan. In the panels *p3* and *p4*, the small blank region around $k_z$ = 12.5 Å$^{-1}$ has not been probed due to the strong intensity signal coming from Cr 3d core levels excited by second-order radiation from the beamline monochromator. The non-dispersive spectral component in Fig. 2 was evaluated by angle integration corrected for angle dependent transmission of the ARPES analyzer. The experimental bands reported in Fig. 3 and 4 ("Exp.") are obtained from MDC (Momentum Distribution Curve) analysis for the three bands crossing the Fermi level, and from EDC (Energy Distribution Curve) analysis for the other two less dispersing bands at binding energies lower than -0.5 eV. The Fermi velocities of the three bands crossing the Fermi level (extracted from MDC analysis) are: (3.8±0.2)×10$^7$ cm/s for the $d_{yz+zx}$ band along the Γ-X direction; (3.7±0.2)×10$^7$ cm/s and (2.6±0.2)×10$^7$ cm/s along the Γ-Z direction for the $d_{yz-zx}$ and $d_{yz+zx}$ bands respectively.

**Theoretical calculations**

The DFT calculations on $CrO_2$ were performed using the Vienna Ab-initio Simulation Package (VASP) [62,63] with a plane-wave cutoff of 350 eV and 6 × 6 × 9 Monkhorst-Pack grid sampling for charge-density integration. The LDA and the GGA-PBE exchange-correlation functionals were used. In both cases, the on-site Coulomb interaction ($U_{eff}$) was tuned to obtain the best agreement with the experimental data, focusing on the most sensitive Γ-Z direction. We adopted the experimental lattice parameters a = b = 4.421 Å, c = 2.917 Å and u = 0.301 [64].

We performed DFT+DMFT calculations [65] using a $t_{2g}$-only model constructed from DFT via maximally localized Wannier functions [66, 67]. The DFT in this case was performed using the VASP package with a plane-wave cutoff of 400 eV and a 16 × 16 × 32 Γ-centered Monkhorst-Pack grid to ensure a faithful representation of the hybridization function also at low temperature. For taking into account the local symmetry of the Cr sites we have used a different local coordinate system on each Cr site. A disentangling procedure with an energy window encompassing the 3d bands of Cr was used, resulting in three $t_{2g}$ orbitals per Cr site, which agree reasonably well in their shapes and transfer integrals with previous calculations [17, 68]. The DMFT part was performed using the code W2DYNAMICS [69] with a continuous-time quantum Monte Carlo impurity solver [70]. We used the density-density Hamiltonian for the $t_{2g}$ orbitals as well as the SU(2)-symmetric Kanamori Hamiltonian. The results in the ordered phase were qualitatively the same in both cases, so only the density-density results are shown. The Coulomb interaction was modeled as usual for a three-band problem with the intraorbital interaction $U$, the Hund's coupling $J$ and the interorbital interaction $U-2J$. The calculations were performed at a temperature of 193 K, well below the ferromagnetic ordering temperature. The spectral functions were obtained by analytical continuation of imaginary time data via the maximum entropy method [71].

We used two different sets of Slater integral parameters: $F^0$ = 1.0eV, $F^2$ = 7.5 eV and $F^4$ = 4.68 eV leading to $U$ = 2.0 eV and $J=J_i$=0.675 eV but $U=F^0$ = 1.0 eV and $J=(F^2+F^4)/14$ = 0.87 eV for the spin-polarized GGA calculations; $F^0$ = 2.0eV, $F^2$ = 7.5 eV and $F^4$ = 4.68 eV corresponding to $U$ = 3.0 eV and $J = J_i$=0.675 eV but $U$ = 2.0 eV and $J$ = 0.87 eV for the spin-degenerate $S_{AVG}$GGA calculations (for the relationship between the different parameters see the following section).

We have also checked different double counting schemes in the GGA+DMFT calculations. In the usual DFT+DMFT the double counting correction (DCC) to the potential is spin independent and can be chosen for example as (Fully Localized Limit) [72]:

$$V_{dc} = U\left(N - \frac{1}{2}\right) - J\left(\frac{N}{2} - \frac{1}{2}\right)$$

In the spin polarized DFT+DMFT one should instead use the spin resolved version of the potential (see e. g. Ref. [73] for a current comparison of different schemes):

$$V_{dc}^{\sigma} = U\left(N - \frac{1}{2}\right) - J\left(N^{\sigma} - \frac{1}{2}\right)$$

In the case of $CrO_2$ the choice only affects the position of the minority spin states, since the position of the majority states is fixed by the requirement of the population by 2 electrons. The results for the occupied bands are not influenced by this choice at all.

**Relationship between Slater and Kanamori parametrizations of the Coulomb interaction**

The Kanamori parameters for the $t_{2g}$ orbitals $U$ and $J$ are connected to the Hubbard parameters for the full *d* shell *U* and *J* as used by, *e.g.*, Liechtenstein *et al.* [40] via the Slater integrals $F^0$, $F^2$, and $F^4$ (for *d* states). Since this relationship is somewhat intricate and not well known in certain communities we outline it below in detail. More can be found in the classical works of Condon and Shortley [74], Slater [75] and Griffith [76].

The tensor of the local Coulomb interaction $U_{mm'm''m'''}$ contains the matrix elements of $e^2/|r-r'|$ in the basis of the wave functions of the atom $\phi_m(r)$, which are composed of a radial part and of a spherical part as usual

$$U_{mm'm''m'''} = \int drdr' \, \phi_m^*(r)\phi_{m'}^*(r') \frac{e^2}{|r-r'|} \phi_{m''}(r)\phi_{m'''}(r')$$

These matrix elements are computed using multipole expansion etc., which is treated in detail in e.g. Ref. [75]. In the course of this calculation the Slater parameters $F^n$ arise as values of the radial parts of the integrals over the wave functions. For the *d* shell only $F^0$, $F^2$, and $F^4$ contribute by symmetry and furthermore we use the approximate relation $F^4=0.625F^2$ [77,78]. These are the only atom dependent parameters remaining, the spherical integrals being universal for the *d* shell, since the spherical parts of the orbitals are given by spherical harmonics.

The tensor can be completely parameterized via the $F^n$. As we do throughout our paper, primarily to facilitate comparison with other DFT+*U* work, a very common notation for the Slater integrals is via the parameters *U* and *J* as follows, see e.g. Ref. [72]

$$U = F^0, \quad J = (F^2 + F^4)/14$$

Once the tensor is computed it can enter the Hamiltonian for the two-article interaction

$$H_{\text{full}} = \frac{1}{2} \sum_{m,m',m'',m'''} \sum_{\sigma\sigma'} U_{mm'm''m'''} d^\dagger_{m\sigma} d^\dagger_{m'\sigma'} d_{m'''\sigma'} d_{m''\sigma}$$

where $d^\dagger_{m\sigma}$ and $d_{m\sigma}$ are the creation and annihilation operators, respectively.

In the DFT+*U* and often also in the DFT+DMFT methodologies an approximation of above Hamiltonian is used, that contains combinations of $d^\dagger_{m\sigma}, d_{m\sigma}$ that can be written as density operators, *i.e.* $n_{m\sigma} = d^\dagger_{m\sigma} d_{m\sigma}$

$$H_{\text{density}} = \frac{1}{2} \sum_{m,m',\sigma} U_{mm'} n_{m,\sigma} \, n_{m,-\sigma} + \frac{1}{2} \sum_{m \neq m',\sigma} (U_{mm'} - J_{mm'}) n_{m,\sigma} \, n_{m',\sigma}$$

In DFT+*U* a Hartree-Fock approximated version of above Hamiltonian enters, where the density operators are approximated by numerical densities.

The direct Coulomb interaction $U_{mm'}$ and exchange $J_{mm'}$ matrices contain only parts of the full tensor and can be explicitly written in terms of the Slater radial integrals as follows. We introduce a few short-hands for brevity

$$U = F^0 + \frac{8}{7}\frac{F^2 + F^4}{14}$$

$$J_1 = \frac{3}{49}F^2 + \frac{20}{9}\frac{1}{49}F^4$$

$$J_2 = -2\frac{5}{7}\frac{F^2 + F^4}{14} + 3J_1$$

$$J_3 = 6\frac{5}{7}\frac{F^2 + F^4}{14} - 5J_1$$

$$J_4 = 4\frac{5}{7}\frac{F^2 + F^4}{14} - 3J_1$$

Now we can write the interaction matrices for a *d* shell, with the real orbitals ordered as $d_{xy}, d_{yz}, d_{3z^2-r^2}, d_{xz}, d_{x^2-y^2}$ in the following compact form, see e.g. Refs. [76, 79, 80, 81] for similar expressions

$$U_{mm'} = \begin{pmatrix} U & U-2J_1 & U-2J_2 & U-2J_1 & U-2J_3 \\ U-2J_1 & U & U-2J_4 & U-2J_1 & U-2J_1 \\ U-2J_2 & U-2J_4 & U & U-2J_4 & U-2J_2 \\ U-2J_1 & U-2J_1 & U-2J_4 & U & U-2J_1 \\ U-2J_3 & U-2J_1 & U-2J_2 & U-2J_1 & U \end{pmatrix}$$

$$J_{mm'} = \begin{pmatrix} 0 & J_1 & J_2 & J_1 & J_3 \\ J_1 & 0 & J_4 & J_1 & J_1 \\ J_2 & J_4 & 0 & J_4 & J_2 \\ J_1 & J_1 & J_4 & 0 & J_1 \\ J_3 & J_1 & J_2 & J_1 & 0 \end{pmatrix}$$

If the off diagonal entries are averaged one obtains the approximate form introduced by Kanamori [81], with the diagonal $U$ and average exchange interaction $J$. Differences between $U$ and $U=F^0$ as well as $J$ and $J=(F^2+F^4)/14$ arise from the fact that $U$ and $J$ are given by the *actual* elements of the Coulomb tensor, whereas *U* and *J* are directly related to the radial Slater integrals. This is a major source of confusion. We note that the diagonal elements of $J_{mm}$ formally exist, but since the corresponding combination of operators would violate the Pauli principle they are put to zero in the matrix. It is clear that the interorbital interactions are different between certain orbitals in the full atomic picture. Thus, there is no single $J$. One also realizes, that the interactions within the $t_{2g}$ orbitals ($d_{xy}$, $d_{yz}$, $d_{xz}$) relevant for CrO$_2$ can be parameterized completely by $U$ and $J_1$. In the manuscript we thus identified $J_1$ as our $J$ for simplicity.

**Unfolding procedure**

The unfolding procedure is obtained by the projections on two basis sets constructed from the $d_{xy}$, $d_{yz-zx}$ and $d_{yz+zx}$ atomic orbitals (for the two different Cr sites in their local coordinate frame) in a way to be even or odd with respect to the screw axis symmetry operator. The wave functions composing the even ("+") or odd ("-") basis set are:

$$\left|\varphi_{xy}^{\pm}\right\rangle = e^{ikR_1}\left|d_{xy}^{Cr1}\right\rangle \pm e^{ikR_2}\left|d_{xy}^{Cr2}\right\rangle$$

$$\left|\varphi_{yz-xz}^{\pm}\right\rangle = e^{ikR_1}\left|d_{yz-xz}^{Cr1}\right\rangle \pm e^{ikR_2}\left|d_{yz-xz}^{Cr2}\right\rangle$$

$$\left|\varphi_{yz+xz}^{\pm}\right\rangle = e^{ikR_1}\left|d_{yz+xz}^{Cr1}\right\rangle \pm e^{ikR_2}\left|d_{yz+xz}^{Cr2}\right\rangle$$

Being composed of only Cr $d$-states, these two basis sets of orbitals (with "+" and "-") do not form a complete one for the system, however, they are sufficient for describing the states near the FS (which is almost completely composed of the Cr $d$-states). For reproducing the matrix elements, the even basis set (composed of the three wave functions with the plus "+") is used for the p-polarization and the odd one ("-") for the s-polarization. As shown in Fig. 5, the projection on the even basis set is complementary to the one on the odd basis set. The partial spectral weight from the $i$-type $d$ atomic orbital associated to the eigenstate $\varepsilon_{n,k}$ of the eigenfunction $\psi_{n,k}$ is calculated as $W_{n,k}^{\pm,i} = \left|\left\langle\varphi_i^{\pm}\middle|\psi_{n,k}\right\rangle\right|^2$, ("+" or "-"depending on even or odd basis set, and so corresponding to p- or s-polarization) and the total spectral

weight is the sum of these three single contributions $W_{n,k}^{\pm,tot} = \sum_i W_{n,k}^{\pm,i}$. The unfolded band structure is represented with the point size proportional to the total spectral weight, while its color is obtained from the percentage representation of colors in the RGB format (red, green and blue) proportional to $W_{n,k}^{\pm,yz+zx}$, $W_{n,k}^{\pm,yz-zx}$ and $W_{n,k}^{\pm,xy}$, respectively.

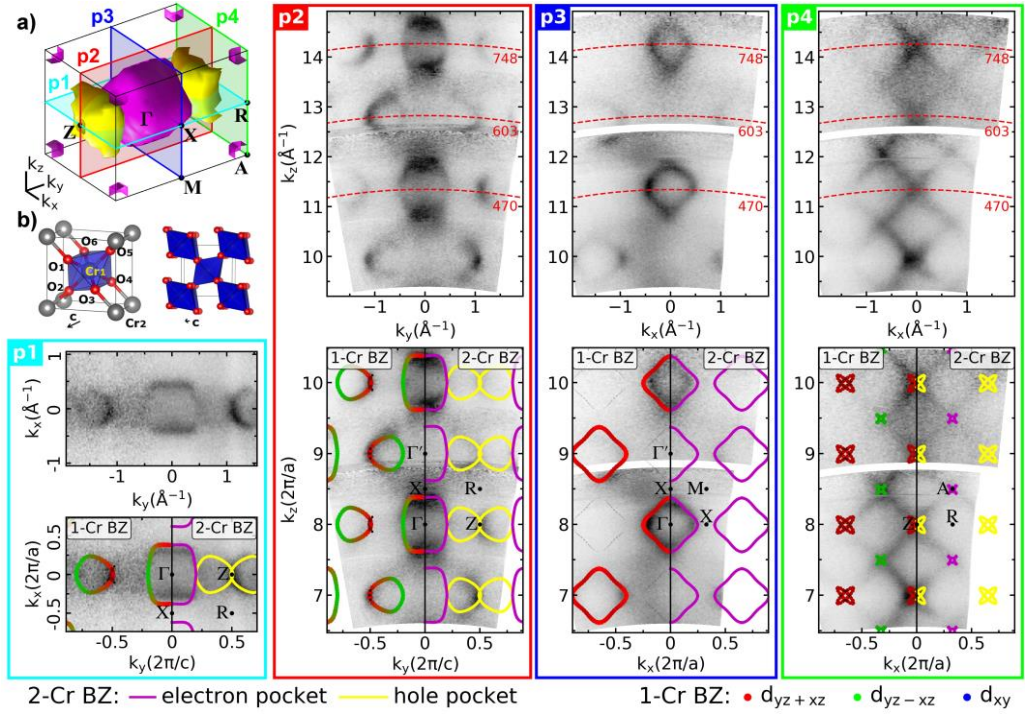

*Fig. 1.* Experimental and theoretical FS of CrO$_2$: (*a*) GGA+*U* ($U_{eff}$ =0.4 eV) calculations; (*b*) Unit cell of CrO$_2$; (*p1-p4*, maps at *top*) Experimental ARPES maps of the FS along the planes marked at (*a*); (*p1-p4*, maps at *bottom*) the same data overlaid with the GGA+*U* ($U_{eff}$ =0.4 eV) results on the right ($k_{x,y}>0$) side. The apparent doubling of the FS periodicity in **k**-space is due to the non-symmorphic space group of CrO$_2$. This is confirmed by unfolding of the calculated FS onto the 1-Cr cell shown on the left ($k_{x,y}<0$) side. These ARPES data were collected with *p*-polarized light at photon energy of 470 eV for *p1* and varied in the 320-820 eV range for (*p2-p4*) as indicated at the top panels.

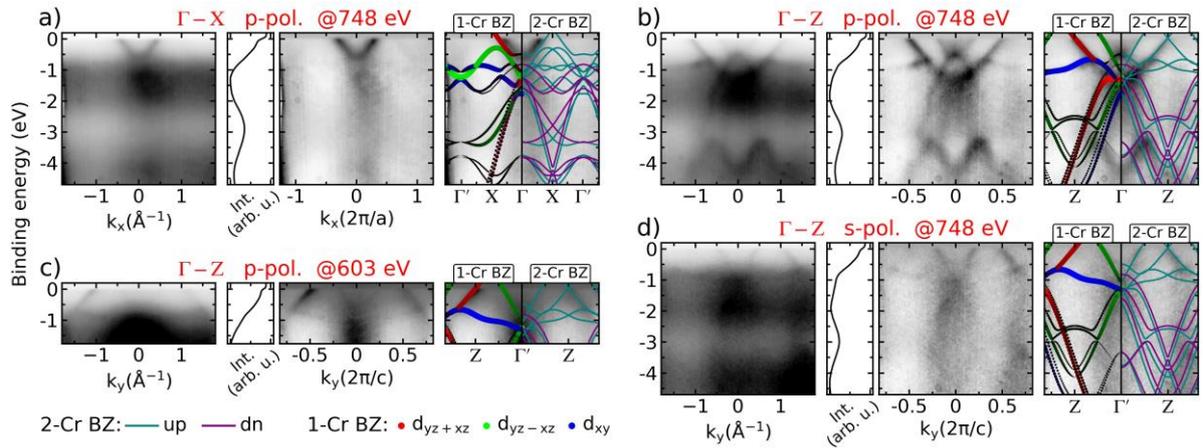

*Fig. 2:* Experimental and theoretical band dispersions in $CrO_2$ along the Γ-X and Γ-Z directions. The photon energy and X-ray polarization are indicated on top of each row *a-d* of the panels representing, from left to right: the raw ARPES image; non-dispersive spectral component coming mostly from the $Cr_2O_3$ overlayer; dispersive spectral component of $CrO_2$; the same data on the $k_{x,y}>0$ side overlaid with the GGA+*U* ($U_{eff}$ = 0.4 eV) dispersions (majority spins in cyan and minority in magenta) and on the $k_{x,y}<0$ side with the corresponding bands unfolded onto the 1-Cr BZ. The unfolding well reproduces the experimental intensity variation through the successive BZ and under X-ray polarization.

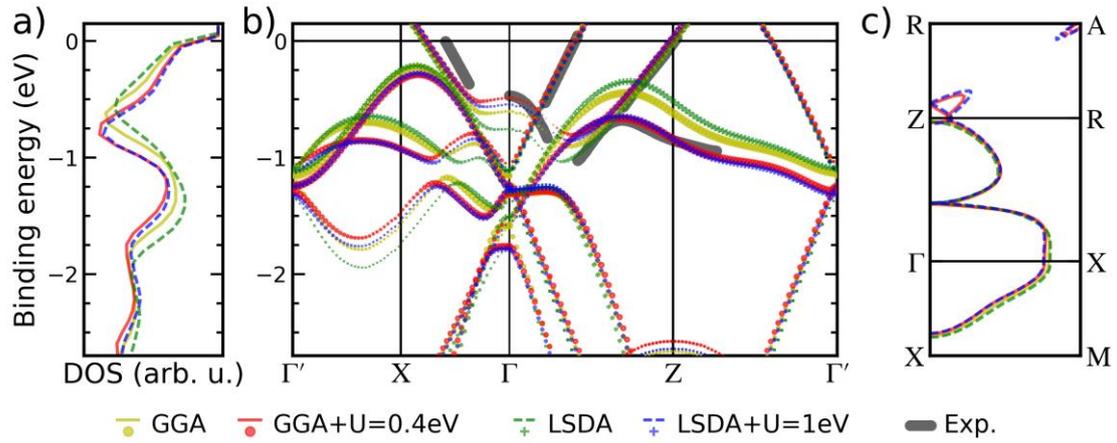

*Fig. 3:* DFT+*U* calculations under variation of $U_{eff}$: (*a*) DOS; (*b*) Unfolded bands along the Γ'-X-Γ-Z-Γ' directions (majority spin only); (c) Corresponding FS. The best match to the experimental bands extracted from ARPES data ("Exp.") is achieved with $U_{eff}$ = 0.4 eV for the GGA and 1 eV for the LSDA exchange-correlation functionals.

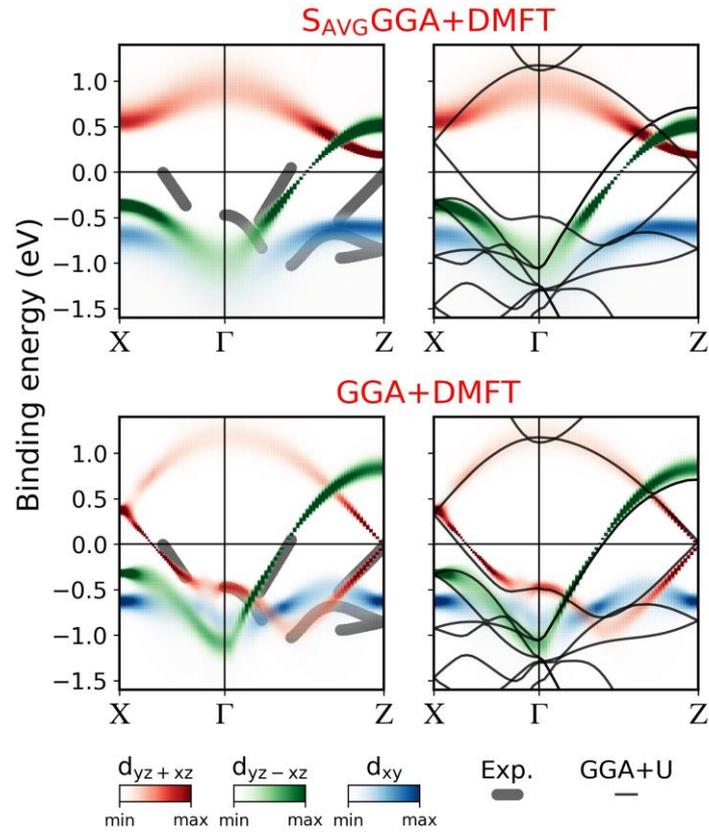

*Fig. 4:* Momentum resolved spectral functions for the majority spin channel within $S_{AVG}$GGA+DMFT (top) and GGA+DMFT (bottom) compared with experimental bands extracted from ARPES (left panels) and GGA+*U* calculations with $U_{eff}$ = 0.4 eV (right panels).

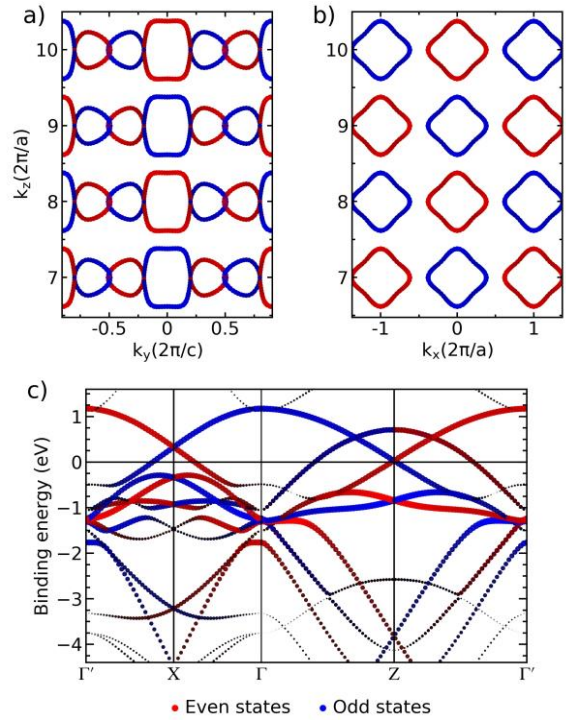

*Fig. 5:* GGA+*U* ($U_{eff}$ = 1 eV) calculations projected on the even (red) and odd (blue) basis set: (a) and (b) FS along the planes p2 and p3, respectively, as reported in *Fig. 1*; (c) bands along the Γ'-X-Γ-Z-Γ' directions (majority spin only). Spots size is proportional to the projection on even and odd basis sets, the color is obtained from the percentage representation of colors in the RGB format proportional to projection on even (odd) basis set for the red (blue) component. The two basis sets complement each other.